# Quantitative Determination of the Band-Gap of $WS_2$ with Ambipolar Ionic Liquid-Gated Transistors

*Daniele Braga[1]\*[†], Ignacio Gutiérrez Lezama[1], Helmuth Berger[2], and Alberto F. Morpurgo[1]\**

[1]DPMC and GAP, Université de Genéve, 24 quai Ernest Ansermet, CH-1211 Geneva, Switzerland

[2]Institut de Physique de la Matiere Complexe, Ecole Polytechnique Federale de Lausanne, CH-1015 Lausanne, Switzerland

We realized ambipolar Field-Effect Transistors by coupling exfoliated thin flakes of tungsten disulphide ($WS_2$) with an ionic liquid-dielectric. The devices show ideal electrical characteristics, including very steep sub-threshold slopes for both electrons and holes and extremely low OFF-state currents. Thanks to these ideal characteristics, we determine with high precision the size of the band-gap of $WS_2$ directly from the gate-voltage dependence of the source-drain current. Our results demonstrate how a careful use of ionic liquid dielectrics offers a powerful strategy to study quantitatively the electronic properties of nano-scale materials.

The discovery that high-quality graphene monolayers - with outstanding optoelectronic properties - can be extracted from graphite crystals by using simple exfoliation techniques *(1)* has prompted research on a broad class of materials produced by similar means *(2)*. Exfoliation works well for most van der Waals layered materials, since the absence of covalent bonds between adjacent layers promotes chemical stability and eliminates problems posed by dangling bonds at the material surface, providing "defect-free" crystals down to the atomic thickness. The investigation of exfoliated crystals is interesting for different reasons. As illustrated by different studies on mono-, bi-, and tri-layer graphene, the properties of layered materials at the atomic scale can differ significantly from those of their bulk parent compounds *(3-5)*. For metallic materials, exfoliated layers can be made as thick as the electrostatic screening length, allowing uniform modulation of their charge density by field-effect *(6)*. The use of thin exfoliated crystals can be advantageous even when their thickness is tens of nanometers (i.e., much thicker than one atom/unit cell), since the small thickness favors better material uniformity and increases the experimental sensitivity to surface properties. This is the case, for instance, of topological insulators, where the use of thin flakes minimizes problems associated with parasitic conduction through the poorly insulating bulk of the material *(7)*.

These considerations apply particularly well to the transition metal dichalcogenides (TMDs), a well-known class of inorganic layered materials in which superconductivity, charge density waves, semi-metallicity, and semiconducting behavior have been observed *(8-13)*. Exfoliated monolayers of molybdenum disulphide ($MoS_2$) were recently found to possess superior optoelectronic properties with respect to their bulk parent compound, as they exhibit a direct semiconducting band-gap - rather than a indirect one *(14,15)* - allowing luminescence *(16)* and enabling the realization of high-performance Field-Effect Transistors (FETs) *(17)*. Thanks to their chemical stability, exfoliated TMDs are also materials of choice to realize ionic liquid-gated

FETs *(18-20)* where an electrolyte-dielectric is employed to accumulate very high densities of surface carriers with the intent to induce, for example, new electronic states of matter. This promising research direction counts few pioneering results showing the possibility to reach surface charge densities in excess of $10^{14}$ cm$^{-2}$ *(21)*, and the occurrence of ambipolar transistor operations with high mobility for both electrons and holes *(22)*. Here we use the same field-effect technique on thin flakes of tungsten disulphide (WS$_2$) to show that ionic liquid-gated FETs, fabricated on high quality exfoliated crystals, enable the quantitative investigation of the material electronic properties.

WS$_2$ is an indirect gap semiconductor ($\Delta_{WS2}$ ~1.35 eV) *(15, 23)* formed by two-dimensional (2D) covalently bonded S-W-S layers separated by a van der Waals gap (Fig. 1 (a)) *(24)*. This semiconductor has been extensively characterized in the past *(25-29)* and bulk crystals have been widely used to realize unipolar FETs working in accumulation of either electrons (n-type) or holes (p-type), depending on the crystal doping and on the preparation technique *(30,31)*. However, the potential of WS$_2$ thin flakes is only now starting to be appreciated *(32)*. We found that, by coupling thin exfoliated flakes of WS$_2$ with an ionic liquid dielectric, stable and hysteresis-free ambipolar transistors with balanced electron and hole conductivities can be realized. The devices characteristics show very steep subthreshold slopes for both electrons and holes –corresponding to threshold voltage swings close to the ultimate room-temperature value of 66 mV/decade– which is indicative of virtually perfect capacitive coupling between the ionic liquid and the WS$_2$ surface, as well as of the high quality of the material. The high material quality is also indicated by the low OFF-state current, which confirm that the density of unintentional dopants and in-gap states in the WS$_2$ thin flake is negligible. Thanks to these ideal transistor characteristics, we show that it is possible to determine with high precision (~10 %) the size of the band-gap of WS$_2$ directly from the gate-voltage dependence of the source-drain current. Next to demonstrating ambipolar transistor operation of unprecedented quality, these results show that a careful use of ionic liquid-gated transistors provides a powerful tool to investigate quantitatively the electronic properties of thin exfoliated crystals.

Thin crystalline flakes of $WS_2$ were obtained by mechanical exfoliation of a bulk single crystal grown by chemical vapor transport and then transferred on a highly doped $Si/SiO_2$ substrate (see Supplementary Information for details on the crystal growth). Flakes suitable for electrical characterization were identified under an optical microscope and their thickness was measured by Atomic Force Microscopy (AFM). Electrical contacts were fabricated by electron-beam lithography, followed by metal (Ti/Au) evaporation and lift-off (the surface of $WS_2$ was Ar-bombarded in situ prior to the metal deposition). We realized approximately 10 different devices on flakes with lateral dimensions ~ 10 µm (depending on the lateral dimensions two- or six-terminal devices were realized; see Fig. 1 (b)), and with thickness ranging from 20 to 60 nm (in the range investigated the device characteristics were found to be thickness independent). The ionic liquid 1-butyl-1-methyl pyrrolidinium tris-(pentafluoroethyl)trifluorophosphate $[P14]^+[FAP]^-$ (Fig. 1 (c)) was chosen as electrolyte-dielectric, because of the wide electrochemical stability window (± 4 V) and of the hydrophobicity of the $[FAP]^-$ anion *(33, 34)* (minimizing the water content in the ionic liquid is essential for the device stability; for the same reason the ionic liquid was placed onto the device in an inert atmosphere and the as-prepared device was rapidly introduced in the vacuum measurement chamber avoiding long-term air exposure). The ionic liquid was retained between the substrate, a gold wire acting as the gate electrode (Fig. 1 (d)) and a rubber well manually deposited onto the substrate. A silver wire treated with piranha solution ($H_2SO_4$ /$H_2O_2$, ratio 3:1) was also inserted in the electrolyte and used as Ag/AgO *quasi*-reference electrode to measure directly the electrochemical potential across the semiconductor/electrolyte interface *(35,36)*. The reference electrode allows us to account for the possible potential drop at the gate/electrolyte interface (Fig. 1 (d); see Supplementary Information for details). All measurements were performed with an Agilent Technology E5270B parameter analyzer at room temperature and in high vacuum ($< 10^{-6}$ mbar), after overnight pumping to minimize the moisture content in the ionic liquid.

Fig. 2 (a) shows the dependence of the source-drain current ($I_D$) of a two-terminal $WS_2$-transistor (measured at source-drain voltage $V_D = 0.1$ V) on the gate voltage $V_G$. The data show ideal, virtually hysteresis-free ambipolar transistor behavior: a large source-drain current $I_D$ is measured at high negative and positive gate voltage - when the Fermi level is deep in the valence and conduction band - and vanishes in between. In the transistor OFF-state - when the Fermi level is in the $WS_2$ band-gap - the current is ~10 pA, corresponding to the experimental sensitivity, indicating a very low density of unintentional dopants in the material, as well as of intra-gap states that could mediate hopping conductivity. Finally, note that this low OFF-state current originates from the high-quality of the $WS_2$ crystals, slowly grown by chemical vapour transport; such a low current is not observed in the characteristics of natural mineral $MoS_2$ crystals that have been used in recent charge transport experiments *(22)*. The ambipolar transistor behaviour is very well-balanced, i.e. at sufficiently high carrier density the same conductance is measured for electron and holes.

Fig. 2 (b) and (c) show log-scale plots of the source-drain current near the onset of hole and electron conduction. A very steep subthreshold slope is visible, corresponding to values of subthreshold swing of $S \sim 90$ meV/decade for electrons and only slightly larger for holes. From the theory of conventional semiconductor FETs, $S = kT/e\ ln(10)(1+ C_D/C_G)$, where $e$ is the electron charge, $C_D$ is the depletion capacitance and $C_G$ the gate capacitance *(37)*. For $C_G >> C_D$, one obtains an ultimate limit of $S = 66$ meV/decade for room-temperature operation, close to the observed experimental value (part of the small observed deviation is due to the effect of the additional capacitance between the gate electrode and the electrolyte). This finding provides a quantitative measure of the excellent electrostatic coupling between the ionic liquid-gate and the FET channel, as well as a clear indication of small charge trapping effects (since the presence of traps would lead to large deviation of *S* from ideality).

The device output characteristics, measured for both polarities of $V_G$ and $V_D$ (Fig. 3 (a)-(b)), confirm the high-quality of the ambipolar transistor operation. Under accumulation of both holes (Fig. 3 (a)) and electrons (Fig. 3 (b)), the current increases at low bias and saturates - as expected - when $V_D$ becomes comparable to $V_G$. Even if a considerable mismatch between the Fermi level of the metal and the $WS_2$ conduction/valence band is present, the $I_D$-$V_D$ curves are linear at low bias as expected for ideal FET characteristics (see Supplementary Information). This is because the injection of charge of both polarities is assisted by electrostatic screening due to ions (in the liquid) close to the metal/semiconductor interface *(38,39)*, which decreases the width of the Schottky barrier down to values comparable to the electrostatic screening length in the ionic liquid ( 1 - 2 nm). As a result, carrier injection from the source/drain contacts occurs via tunneling under the Schottky barrier for both electrons and holes, and not by an over-the-barrier thermal activation process (tunnelling-mediated carrier injection is likely also the reason for the steep subthreshold slope that is observed independently of the device contact resistance). Finally, the steep increase in $I_D$ at large (absolute) $V_D$ and small $V_G$ visible in Fig. 3 (a) and (b) is another clear manifestation of ambipolar operations. When $V_D$ (in absolute value) is increased well beyond $V_G$, the polarity of the channel potential relative to the gate is opposite at the source and drain electrode, resulting in the simultaneous injection of electrons and holes at the two different contacts *(40)*. This transport regime, in which the current is carried by either electrons or holes in different parts of the channel, is commonly observed in high-quality ambipolar transistors *(41)* and has been studied extensively to realize, for example, the so-called organic light-emitting transistors *(42,43)*.

More information about the semiconductor characteristics can be obtained by analyzing quantitatively the $V_G$-range corresponding to the device OFF-state, where the Fermi level is swept across the gap of WS$_2$. The extension of this range corresponds to the difference between the threshold voltages for electron and hole conduction (i.e., $\Delta V_{GAP} = (V^e_{TH} - V^h_{TH})$). It is apparent from Fig. 2 (a) that $\Delta V_{GAP} \sim 1.4$ V, but to make the analysis more precise we looked at the data as a function of the reference voltage $V_{REF}$, and measured the transfer curves for different $V_D$ to obtain the values of $V^e_{TH}$ and $V^h_{TH}$ by extrapolating to $V_D = 0$ V. The procedure is illustrated in Fig. 4 (a)-(e): we find $\Delta V_{GAP} = (V^e_{TH} - V^h_{TH}) = 1.4$ V, with a ~ +/- 5% experimental error originating from the extrapolation procedure (see Fig. 4 (e)). This value nearly perfectly coincides with the known band-gap of WS$_2$ (1.3- 1.35 eV) *(15,23)*. As an independent confirmation of this result we also measured the displacement current flowing through the gate ($I_G$) *(44)*, while sweeping $V_G$ (Fig. 4 (b)) with both source and drain electrodes grounded ($V_S = V_D = 0$ V) *(45)*. This current is mainly associated to charging of the capacitors at both electrolyte/gate and electrolyte/semiconductor interfaces. The peaks in $I_G$ at approximately $V_{REF} \sim -0.9$ V and $V_{REF} \sim +0.4$ V correspond to a capacitance change associated to accumulation of holes and electrons at the surface of WS$_2$. The distance between the peaks is approximately 1.25 V, very close to the value of $\Delta V_{GAP} = 1.4$ V extracted from the FET characteristics. Considering the ultimate energy resolution at room temperature (approximately 3.5 kT ~ 90 meV), the two experimental values are consistent with each other and with the band-gap of the material with a 10% precision. We conclude, therefore, that we can precisely determine the band-gap of WS$_2$ from simple transport measurements on a nano-fabricated FET.

This conclusion is remarkable because extracting the band-gap of a semiconductor from the transfer characteristics of a conventional transistor ($I_D$ versus $V_G$) is usually not possible for several reasons. These include the contact effects (causing large and uncontrolled threshold voltage shifts, not present in our devices where charge injection is mediated by tunneling through the Schotkky barrier –as we explained above), the influence of localized energy gap-states on charge transport (which provide a parallel conduction path preventing the precise determination of the transistor OFF-state, and affect the relation between the change in $V_G$ and the corresponding change in the chemical potential, see below), and the rather small capacitance of conventional solid-state dielectrics. The relevance of the last two phenomena can be understood by recalling that the change in the FET gate bias ($V_G$) is related to the change of the semiconductor Fermi level position by *(46)*:

$$e\Delta V_G = \Delta E_F + e\Delta\phi = \Delta E_F + \frac{e^2 n}{C_G}, \qquad (1)$$

i.e., a change in $V_G$ equals the change of the material chemical potential $\Delta E_F$ plus the variation of the electrostatic potential $e\Delta\phi$, which for a FET is given by $\frac{e^2 n}{C_G}$ ($n$ is the surface density of accumulated charge; Eq. (1) assumes that the FET channel is sufficiently long to disregard the capacitance between the source/drain electrodes and the channel, as in our devices). If Eq. (1) is applied to the $V_G$ interval corresponding to the transistor OFF-state, $e\Delta V_{GAP} = \Delta E_F = \Delta_{WS2}$ only if the term $\frac{e^2 n}{C_G}$ can be neglected. In general this cannot be done: this term usually dominates because there is a finite density of states in the band gap due to defects in the material (i.e., charge is accumulated in localized states inside the band-gap) and because in common experiments with solid-state gate dielectrics the capacitance $C_G$ is relatively small. However, the ionic liquid-gate makes $C_G$ extremely large, and our WS$_2$ crystals have a negligibly small density of in-gap states, as shown by both the ideal value of the subthreshold swing S and the low current in the OFF-state. As a result, the term $\frac{e^2 n}{C_G}$ is negligible and $e\Delta V_{GAP} = \Delta_{WS2}$, as found experimentally.

Finally, we discuss the Hall-Effect measurements that we performed to determine the mobility and the density of the charge carriers. As it is apparent from Fig. 5 (a), the transverse resistance $R_{xy}(B)$ is linearly dependent on the magnetic field $B$ at all different positive and negative $V_G$ values, exhibiting a sign-change when carriers of different polarity are accumulated in the channel. The inverse of the slope of $R_{xy}(B)$ gives directly the carrier density $n_{e,h}$, and the corresponding mobility ($\mu_{e,h} = \sigma_{e,h}/en_{e,h}$) can be calculated from the channel longitudinal conductivity $\sigma_{e,h}$ (that is measured simultaneously). In Fig. 5 (b) and (c) both quantities are plotted for electrons and holes as a function of $V_G$ (or, more precisely, as a function of $V_G$-$V^{e,h}_{TH}$, i.e. the gate voltage relative to the corresponding threshold voltage). At a comparable distance from the threshold, the electron density is approximately three times larger than the hole density (at the maximum $V_G$ reached in the measurements $n_e \sim 10^{14}$ cm$^{-2}$ and $n_h \sim 3.5\ 10^{13}$ cm$^{-2}$). At the same time, at large gate voltages holes have larger mobility than electrons, so that the conductivities of electron and holes are comparable (at large $V_G$, $\mu_e \sim 20$ cm$^2$/Vs and $\mu_h \sim 90$ cm$^2$/Vs). An asymmetry between electron and hole density/mobility in ionic liquid-gated transistors could originate from the different size of anions and cations (the different size leads to different gate capacitance and different ion/carrier interaction) *(45)*. However, the difference between the density of states (DOS) of the WS$_2$ valence and conduction bands is also likely to give an important contribution to the asymmetry. Theoretical calculations indicate that the width of the valence band in WS$_2$ is approximately twice as large as that of the conduction band, implying that the DOS is correspondingly two times smaller in the valence as compared to the conduction band *(15)*. Owing to the very large capacitance of the ionic liquid, the amount of carriers accumulated in the semiconductor is limited by the DOS, because of the so-called quantum capacitance (see for instance the comparison between mono-, bi-, and tri-layer graphene in Ref. *(46)*). Therefore, a smaller DOS in the valence band leads directly to a smaller density of accumulated charge. Moreover, the difference in the density of states can also account for the difference in mobility, since within a simple Born approximation the scattering rate is proportional to the DOS (this mechanism could explain, therefore, the balanced conductivities observed in the experiments).

In conclusion, ionic liquid-gated FETs made on $WS_2$ thin flakes are proven to be ideal ambipolar devices thanks to a virtually perfect electrostatic coupling between the ionic liquid-gate and the transistor channel, and to the very high quality of the employed semiconductor. Our results, together with their analysis, show that ionic liquid-gated transistors offer new strategies to study quantitatively the electronic properties of nano-materials. In particular, the technique discussed here is well-suited for the investigation of layers whose thickness can be reduced down to the atomic scale.

**Corresponding Authors**

* (D. B) braga@ipe.mavt.ethz.ch. * (A. F. M.) alberto.morpurgo@unige.ch

**Present Addresses**

†Optical Materials Engineering Laboratory, ETH Zurich, Universitätstrasse 6, 8092 Zürich, Switzerland.

**Acknowledgment**

We gratefully acknowledge A. Ferreira, N. Couto, and N. Minder for technical support and help, as well as Prof. László Forró for his support. Financial support from the Swiss National Science Foundation and from the NCCR MaNEP is also acknowledged. The $WS_2$ crystal structure in Fig. 1 (a) has been processed by the free-software Mercury *(47)*.

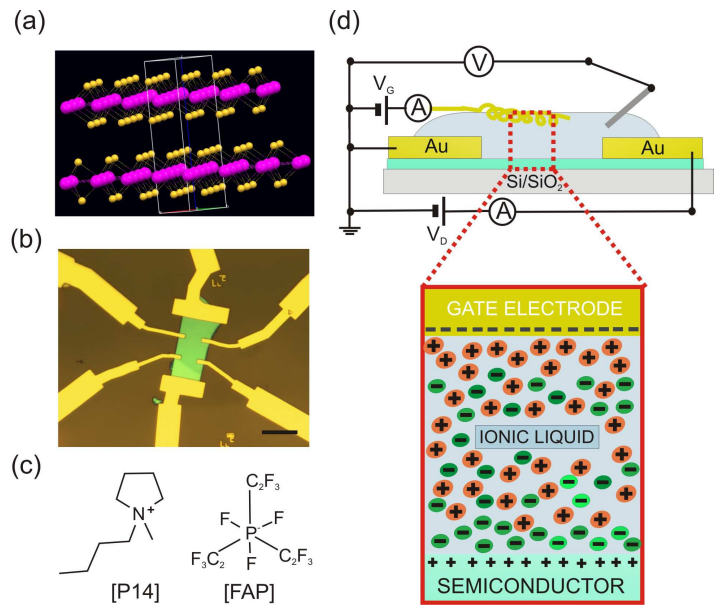

Figure 1. (a) Crystal structure of layered WS$_2$; the purple balls correspond to W atoms, the yellow balls to S atoms. (b) Optical microscope image of a six-terminal Hall-bar fabricated on a crystalline WS$_2$ thin flake (the scale bar is 10 µm long). (c) Molecular structure of the ionic liquid employed as electrolyte-gate dielectric. Both cation [P14]$^+$ and anion [FAP]$^-$ are shown. (d) Cross section of a WS$_2$ ionic liquid-gated FET. A large area gold mesh and an oxidized Ag/AgO wire are used as gate electrode and *quasi*-reference electrode, respectively. When the gate electrode is biased, two electrical double layers are formed at the gate/electrolyte and semiconductor/electrolyte interfaces, enabling charge accumulation at the semiconductor surface.

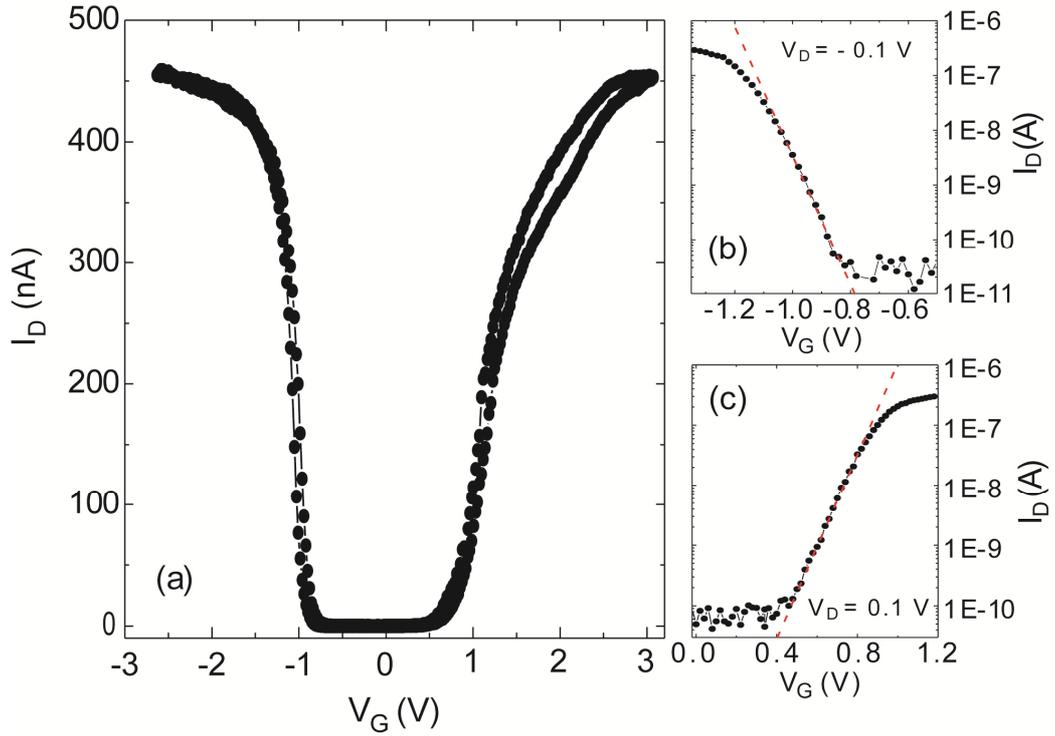

Figure 2. (a) Transfer characteristic of a $WS_2$ ionic liquid-gated FET measured at $V_D = 0.1$ V, as a function of gate voltage $V_G$ (sweep rate 10 mV/s). (b)-(c) ($I_D$-$V_G$) characteristics at the onset of hole and electron conduction, with $I_D$ plotted in logarithmic scale to show the steep subthreshold slopes observed for both polarities. The dashed lines represent best fits of the subthreshold slope for holes (b) and electrons (c).

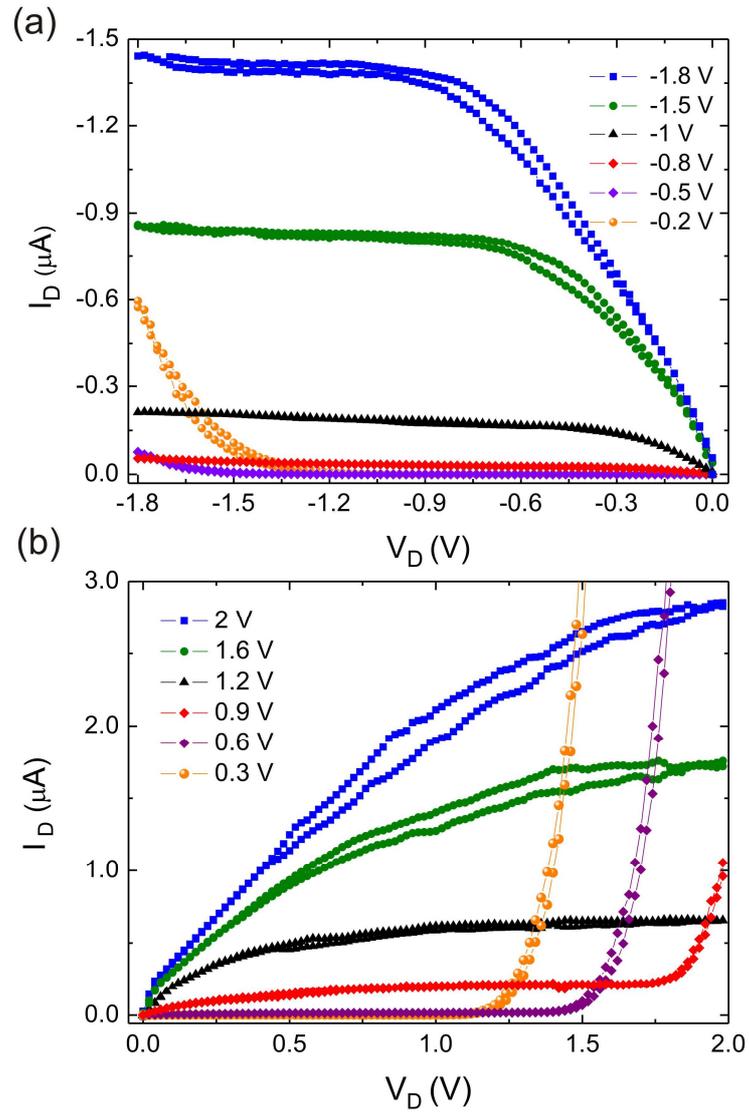

Figure 3. Drain current ($I_D$) versus drain voltage ($V_D$) at different negative (a) and positive (b) gate voltages ($V_G$) (as indicated in the legends of the corresponding panels). The drain current increases linearly at low bias and saturates when $V_D$ becomes comparable to $V_G$, as for a conventional MOSFET. Clear ambipolar charge transport manifests itself at low $V_G$ and at sufficiently large $V_D$ of the same polarity, through a steep current increase due to simultaneous injection of electrons and holes from the two contacts.

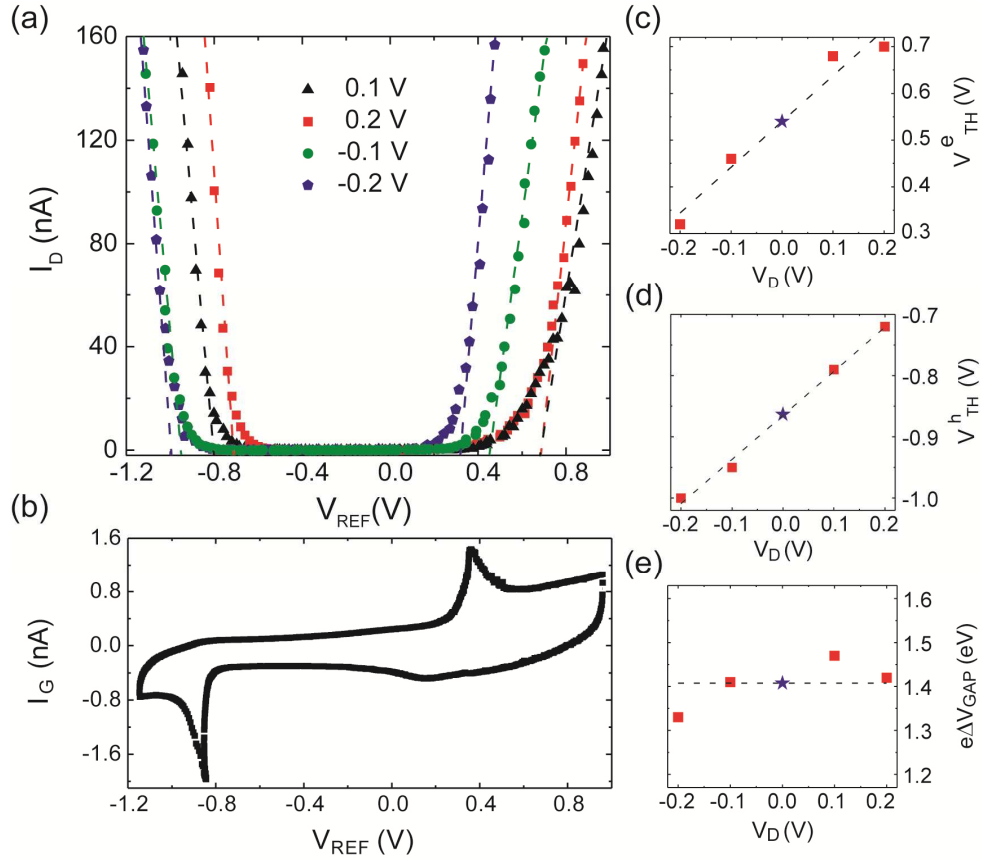

Figure 4. (a) Drain current ($I_D$) versus reference voltage ($V_{REF}$) at different drain voltages ($V_D$). (b) Gate electrode current ($I_G$) versus ($V_{REF}$) (transistor displacement current). $I_G$ is measured between the gate electrode and the source electrode while sweeping $V_G$. Both the source and the drain electrode are grounded to avoid variation of the longitudinal electrostatic potential in the channel. (c-d) Onset of electron ($V^e_{TH}$) and hole ($V^e_{TH}$) conduction (threshold voltage) versus drain voltage ($V_D$); the threshold voltage values at different $V_D$ have been determined by linearly extrapolating to zero the ($I_D - V_G$) characteristics, as indicated by the dashed lines in Fig. 4 (a). (e) WS$_2$ energy band-gap calculated from the threshold voltages reported in Fig. 4 (c-d): $e\Delta V_{GAP} = e(V^e_{TH} - V^h_{TH})$.

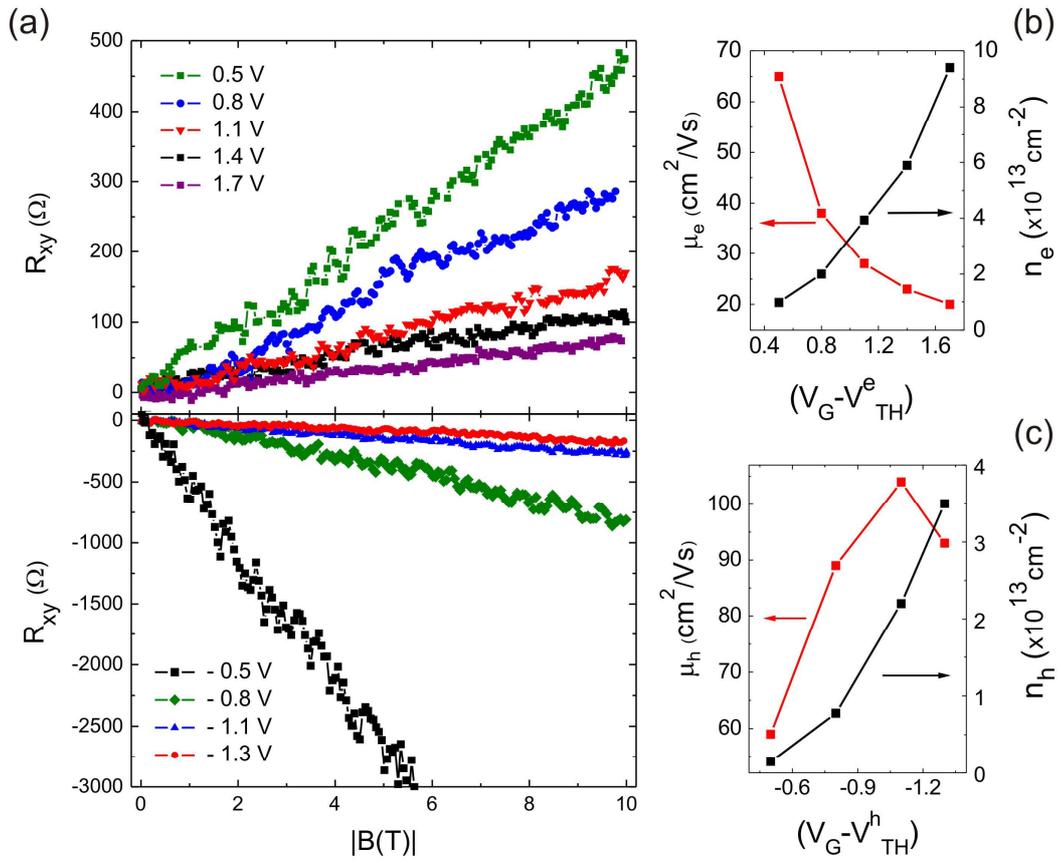

Figure 5. Hall-effect in WS$_2$ ionic liquid gated FETs. (a) Transverse resistance $R_{xy}(B)$ versus $B$ for different positive (upper panel) and negative (lower panel) gate biases ($V_G - V^{e,h}_{TH}$) (as indicated in the corresponding legends). Electron (b) and hole (c) carrier density ($n_{e-h}$) and Hall mobility ($\mu_{e-h}$), as a function of ($V_G - V^{e,h}_{TH}$).

**References**

(1) Geim, A. K.; Novoselov, K. S. *Nat. Mater.* **2007**, *6*, 183.

(2) Coleman, J. N.; Lotya, M.; O'Neill, A.; Bergin, S. D.; King, P. J.; Khan, U.; Young, K.; Gaucher, A.; De, S.; Smith, R. J.; Shvets, I. V.; Arora, S. K.; Stanton, G.; Kim, H.-Y.; Lee, K.;Kim, G. T.; Duesberg, G. S.; Hallam, T.; Boland, J. J.; Wang, J. J.; Donegan, J. F.; Grunlan, J. C.; Moriarty, G.; Shmeliov, A.; Nicholls, R. J.; Perkins, J. M.; Grieveson, E. M.; Theuwissen, K.; McComb, D. W.; Nellist, P. D.; Nicolosi, V. *Science* **2011**, *331*, 568.

(3) Castro Neto, A. H.; Guinea, F.; Peres, N. M. R.; Novoselov, K. S.; Geim, A. K. *Rev. Mod. Phys.* **2009**, *81*, 109.

(4) Oostinga, J. B.; Heersche, H. B.; Liu, X.; Morpurgo, A. F.; Vandersypen, L. M. K. *Nat. Mater.* **2007**, *7*, 151.

(5) Craciun, M. F.; Russo, M. F.; Yamamoto, M.; Oostinga, J. B.; Morpurgo, A. F.; Taruca, S. *Nat. Nanotech.* **2009**, *4*, 383.

(6) Novoselov, K. S.; Jiang, D.; Schedin, F.; Booth, T. J.; Khotkevich, V. V.; Morozov, S. V.; Geim, A. K. *Proc. Natl. Acad. Sci.* **2005**, *102*, 10451.

(7) Sacépé, B.; Oostinga, J. B.; Li, J.; Ubaldini, A.; Couto, N. J. G.; Giannini, E.; Morpurgo, A. F. *Nat. Commun*. **2011**, *2*, 575.

(8) Marseglia, E. A. *Int. Rev. Phys. Chem.* **1983**, *3*, 177.

(9) Morosan, E.; Zandbergen, H. W.; Dennis, B. S.; Bos, J. W. G.; Onose, Y.; Klimczuk, T.; Ramirez, A. P.; Ong, N. P.; Cava, R. J. *Nat. Phys.* **2006**, *2*, 544.

(10) Shen, D. W.; Xie, B. P.; Zhao, J. F.; Yang, L. X.; Fang, L.; Shi, J.; He, R. H.; Lu, D. H.; Wen, H. H.; Feng, D. L. *Phys. Rev. Lett.* **2007**, *99*, 216404.

(11) Rossnagel, K. *J. Phys. Cond. Mat.* **2011**, *23*, 213001.

(12) Podzorov, V.; Gershenson, M. E.; Kloc, Ch.; Zeis, R.; Bucher, E. *Appl. Phys. Lett.* **2004**, *84*, 3301.

(13) Ayari, A.; Cobas, E.; Ogundadegbe, O.; Fuhrer, M. S. *J. Appl. Phys.* **2007**, *101*, 014507.

(14) Ramakrishna Matte, H.S.S.; Gomathi, A.; Manna, A. K.; Late, D. J.; Datta, R.; Pati, S. K.; Rao, C. N. R. *Angew. Chem. Int. Ed.* **2010**, *49*, 4059.

(15) Kuc, A.; Zibouche, N.; Heine, T. *Phys. Rev. B.* **2011**, *83*, 245213.

(16) Splendiani, A.; Sun, L.; Zhang, Y.; Li, T.; Kim, J.; Chim, C.-Y.; Galli, G.; Wang, F. *Nano Lett.* **2010**, *10*, 1271.

(17) Radisavljevic, B.; Radenovic, A.; Brivio, J.; Giacometti, V.; Kis, A. *Nat. Nanotech.* **2011**, *6*, 147.

(18) Panzer, M. J.; Newman C. R.; Frisbie, C. D. *Appl. Phys. Lett.* **2005**, 86, 103503.

(19) Shimotani, H; Asanuma, H.; Takeya, J.; Iwasa, Y. *Appl. Phys. Lett.* **2006**, 89, 203501.

(20) Ueno, K.; Nakamura, S.; Shimotani, H.; Ohtomo, A.; Kimura, N.; Nojima, T.; Aoki, H.; Iwasa, Y.; Kawasaki, M. *Nat. Mater.* **2008**, 7, 855.

(21) Yuan, H. T.; Toh, M.; Morimoto, K.; Tan, W.; Wei, F.; Shimotani, H.; Kloc, Ch.; Iwasa, Y. *Appl. Phys. Lett.* **2011**, *98*, 012102.

(22) Zhang, Y.; Ye, J.; Matsuhashi, Y.; Iwasa, Y. *Nano Lett.* **2012**, *12*, 1136.

(23) Kam, K. K.; Parkinson, B. A. *J. Phys. Chem.* **1982**, *86*, 463.

(24) Schutte, W. J.; de Boer, J. L.; Jellinek, F. *J. Sol. St. Chem.* **1987**, *70*, 207.

(25) Baglio, J. A.; Calabrese, G. S.; Kamieniecki, E.; Kershaw, R.; Kubiak, C. P.; Ricco, A. J.; Wold, A.; Wrighton, M. S.; Zoski, G. D. *J. Electrochem. Soc.* **1982**, *129*, 1461.

(26) Baglio, J. A.; Calabrese, G. S.; Harrison, D. J.; Kamieniecki, E.; Ricco, A. J.; Wrighton, M. S.; Zoski, G. D. *J. Am. Chem. Soc.* **1983**, *105*, 2246.

(27) Ohuchi, F. S.; Jaegermann, W.; Pettenkofer, C.; Parkinson, B. A. *Langmuir* **1989**, *5*, 439.

(28) Genut, M.; Margulis, L.; Tenne, R.; Hodes, G. *Thin Solid Films* **1992**, *219*, 30.

(29) Ki, W.; Huang, X.-Y.; Li, J.; Young, D. L.; Zhang, Y. *J. Mater. Res.* **2007**, *22*, 1390.

(30) Baglio, J. A.; Kamieniecki, E.; DeCola, N.; Struck, C. *J. Sol. St. Chem.* **1983**, *49*, 166.

(31) Matthaus, A.; Ennaoui, A.; Fiechter, S.; Tiefenbacher, S.; Kiesewetter, T.; Diesner, K.; Sieber, I.; Jaegermann, W.; Tsirlina, T.; Tenne, R. *J. Electrochem. Soc.* **1997**, *144*, 1013.

(32) Hwang, W. S.; Remskar, M.; Yan, R.; Protasenko, V.; Tahy, K.; Chae, S. D.; Zhano, P.; Konar, A.; Xing, H. G.; Seabaugh, A.; Jena, D. *Appl. Phys. Lett.* **2012**, *101*, 013107.

(33) Paulsen, B. D.; Frisbie, C. D. *J. Phys. Chem. C* **2012**, *116*, 3132.

(34) O' Mahony, A. M.; Silvester, D. S.; Aldous, L.; Hardacre, C.; Compton, R. G. *J. Chem. Eng. Data* **2008**, *53*, 2884.

(35) Ofer, D.; Crooks, R. M.; Wrighton, M. S. *J. Am. Chem. Soc* **1990**, *112*, 7869.

(36) Xia, Y.; Cho, J.; Paulsen, B.; Frisbie, C. D.; Renn, M. J. *Appl. Phys. Lett.* **2009**, *94*, 013304.

(37) Sze, S. M. *Physics of Semiconductor Devices*, 2nd ed.; John Wiley, New York, (**1981**).

(38) Shimotani, H.; Asanuma, H.; Iwasa, Y. *Jpn. J. Appl. Phys.* **2007**, *46*, 3613.

(39) Braga, D.; Ha, M.; Xie, W.; Frisbie, C. D. *Appl. Phys. Lett.* **2010**, *97*, 193311.

(40) Neudeck, G. W.; Bare, H. F.; Chung, K. Y. *IEEE Trans. Electron Dev.* **1987**, *34*, 344.

(41) Meijer, E. J.; de Leeuw, D. M.; Setayesh, S.; van Veenendaal, E.; Huisman, B. H.; Blom, P. W. M.; Hummelen, J. C.; Scherf, U.; Klapwijk, T. M. *Nat. Mater.* **2003**, *2*, 678.

(42) Capelli, R.; Toffanin, S.; Generali, G.; Usta, H.; Facchetti, A.; Muccini, M. *Nat. Mater.* **2010**, *9*, 496.

(43) It would be interesting to investigate the evolution of transport in this regime as a function of the $WS_2$ thickness, since $WS_2$ monolayers (as the case of $MoS_2$) are predicted to possess a direct band-gap, and therefore to allow electrically-driven emission of light (which is not the case for the thicker crystals used in this study). In this context, we note that an important goal of organic electronics is the realization of electrically controlled lasing using light emitting transistors. Despite impressive progress, a remaining obstacle with organic semiconductors is that their rather low carrier mobility limits the recombination rate of electron-hole pairs, and that high-mobility organic single-crystals supporting ambipolar transport are still sought for the device fabrication. The high charge carrier mobility and the well-balanced electron and hole conductivity in monolayer of TMDs, such as $MoS_2$ or $WS_2$, make these materials suitable for the realization of FET lasers.

(44) Liang, Y. ; Chang, H.; Ruden, P. P.; Frisbie, C. D. *J. Appl. Phys.* **2011**, 110, 064514.

(45) Xie, W.; Frisbie, C. D. *J. Phys. Chem. C* **2011**, *115*, 14360.

(46) Ye, J.; Craciun, M. F.; Koshino, M.; Russo, S.; Inoue, S.; Yuan, H.; Shimotani, H.; Morpurgo, A. F.; Iwasa, Y. *Proc. Natl. Acad. Sci.* **2011**, *108*, 13002.

(47) Macrae, C.F.; Bruno, I. J. ; Chisholm, J. A.; Edgington, P. R.; McCabe, P.; Pidcock, E.; Rodriguez-Monge, L.; Taylor, R.; van de Streek, J.; Wood, P.A *J. Appl. Cryst.* **2008**, *41*, 466.